\documentclass{article}
\usepackage[margin=1in]{geometry}
\usepackage[onehalfspacing]{setspace}
\usepackage{amsmath,mathtools}
\usepackage{pgfplots}
\usepackage{subcaption}
\usepackage{booktabs}
\usepackage{graphicx}
\usepackage{authblk}
\usepackage{xcolor}
\usepackage{booktabs}
\usepackage{subcaption}
\usepackage[linesnumbered,ruled,vlined]{algorithm2e}
\usepackage{qcircuit}
\usepackage{pgffor}
\usepackage{ifthen}
\usepackage{tcolorbox}

\SetCommentSty{mycommfont}

\SetKwInput{KwInput}{Input}                
\SetKwInput{KwOutput}{Output}              
\SetKwProg{Fn}{Function}{}{}

\usepackage{xinttools}
\usepackage{xintexpr}
\usepackage{multicol}
\usepackage{centeredline}
\usepackage{nicematrix}
\usepackage{arydshln}

\newcommand{\lob}[1]{\left( #1 \right)} 

\usepackage[english]{babel}

\usepackage{pdflscape}
\usepackage{fancyvrb}
\usepackage{calc}
\usepackage{rotating}
\usepackage{pgf,tikz}
\usepackage{mathrsfs}
\usetikzlibrary{arrows}
\usepackage{mathtools}

\usepackage{listings}
\usepackage[]{qcircuit}
\usepackage{mathtools}
\usepackage{varwidth}
\usepackage{caption}
\usepackage{subcaption}
\usepackage{graphicx}
\usepackage[export]{adjustbox}

\usepackage{float}

\usepackage{mathtools}
\DeclarePairedDelimiter\bra{\langle}{\rvert}
\DeclarePairedDelimiter\ket{\lvert}{\rangle}
\DeclarePairedDelimiterX\braket[2]{\langle}{\rangle}{#1\,\delimsize\vert\,\mathopen{}#2}

\makeatletter
\def\paragraph{\@startsection{paragraph}{4}%
	\z@\z@{-\fontdimen2\font}%
	{\normalfont\bfseries}}
\makeatother

\usepackage{graphicx}
\usepackage{pgffor}
\usepackage{tikz,tikz-cd}
\usetikzlibrary{backgrounds,fit,decorations.pathreplacing}  
\usepackage{booktabs}
\usepackage{enumerate}

\usepackage{amssymb, amsmath, amsthm}

\usepackage{xypic}
\usepackage{scalerel}
\usepackage{ulem}

\usepackage{graphicx}              
\usepackage{amsmath}               
\usepackage{amsfonts}              
\usepackage{amsthm}                
\usepackage{xkeyval}               
\usepackage{amssymb}
\usepackage{enumerate}             
\usepackage{color}
\usepackage{breqn}                 
\usepackage{bm}                    
\usepackage{cite}

\allowdisplaybreaks[1]
\usepackage{nicefrac}
\usepackage{booktabs}

\usepackage{comment}

\usepackage{kpfonts}
\usepackage{stackengine}
\usepackage{calc}
\newlength\shlength
\newcommand\xshlongvec[2][0]{\setlength\shlength{#1pt}%
	\stackengine{-5.6pt}{$#2$}{\smash{$\kern\shlength%
			\stackengine{7.55pt}{$\mathchar"017E$}%
			{\rule{\widthof{$#2$}}{.57pt}\kern.4pt}{O}{r}{F}{F}{L}\kern-\shlength$}}%
	{O}{c}{F}{T}{S}}

\usepackage{nicematrix}

\usepackage{hyperref}
\hypersetup{
	colorlinks   = true,    
	citecolor    = red      
}

\newcommand{\RN}[1]{%
	\textup{\uppercase\expandafter{\romannumeral#1}}%
}

\allowdisplaybreaks

\newcommand{\meqref}[1]{\text{Eq}.~\eqref{#1}}
\newcommand{\mref}[1]{Sec.~$ \!\ref{#1} $}
\newcommand{\mfig}[1]{Fig.~$ \!\ref{#1} $}

\newtheorem{thm}{Theorem}[section]

\newtheorem{remark}[thm]{Remark}

\newcommand{\mat}[4]{\left[\begin{smallmatrix*}[r]
		#1 & #2 \\
		#3 & #4 \\
	\end{smallmatrix*}\right]}

\def\<{\langle}
\def\>{\rangle}

\numberwithin{equation}{section}

\usetikzlibrary{matrix,fit}

\usepackage{pgfplots}
\pgfplotsset{compat=1.17}

\makeatletter
\def\smallunderbrace#1{\mathop{\vtop{\m@th\ialign{##\crcr
				$\hfil\displaystyle{#1}\hfil$\crcr
				\noalign{\kern3\p@\nointerlineskip}%
				\tiny\upbracefill\crcr\noalign{\kern3\p@}}}}\limits}
\makeatother

\begin{document}
	\title{Efficient quantum algorithm for weighted partial sums and numerical integration} 
	\author[1]{Alok Shukla \thanks{Corresponding author.}}
	\author[2]{Prakash Vedula}
	\affil[1]{School of Arts and Sciences, Ahmedabad University, India}
	\affil[1]{alok.shukla@ahduni.edu.in}
	\affil[2]{School of Aerospace and Mechanical Engineering, University of Oklahoma, USA}
	\affil[2]{pvedula@ou.edu}
	
\date{}

	\maketitle

\begin{abstract}
This paper presents a quantum algorithm for efficiently computing partial sums and specific weighted partial sums of quantum state amplitudes. 
Computation of partial sums has important applications, including numerical integration, cumulative probability distributions, and probabilistic modeling. The proposed quantum algorithm uses a custom unitary construction to achieve the desired partial sums with gate complexity and circuit depth of \(O(\log_2 M)\), where \(M\) represents the number of terms in the partial sum. 
For cases where \(M\) is a power of two, the unitary construction is straightforward; however, for arbitrary \(M\), we develop an efficient quantum algorithm to create the required unitary matrix. Computational examples for evaluation certain partial sums and numerical integration based on our proposed algorithm are provided.
We also extend the algorithm to evaluate partial sums of even or odd components and more complex weighted sums over specified intervals. 
\end{abstract}

\section{Introduction}\label{sec:intro}
We present a quantum algorithm for efficiently computing partial sums, including specific weighted partial sums of quantum state amplitudes, with potential applications in numerical integration and related computational problems.
The goal is to compute the partial sum
\begin{align} \label{eq_partial_sum}
    S_M = \sum_{k=0}^{M-1} f_k, 
\end{align}
for any given $M$, with $ 1 < M \leq N = 2^n$, where
 \( {\bm{f}} = \begin{bmatrix} f_0 & f_1 & \hdots & f_{N-1} \end{bmatrix}\) is the given normalized input vector, i.e., $\sum_{k=0}^{N-1} \left|f_k \right|^2 =1$.  
This problem is directly related to numerical integration, where partial sums are used as a discrete approximation to integrals of functions defined over intervals. For instance, integrating a function \( f(x) \) over an interval \([0,a]\) can be approached by discretizing the interval into \( N = 2^n \) points and summing up the function values at these points, analogous to the computation of partial sums. The difference is that in the case of integration, the values are multiplied by the interval width before summation, which effectively applies a weight to each function value.

The computation of partial sums is also critical in determining cumulative probability distributions. When the vector \( \begin{bmatrix} f_0 & f_1 & \hdots & f_{N-1} \end{bmatrix}\) represents the probability mass function (PMF) of a discrete random variable, the partial sum \( S_M \) gives the cumulative probability up to the \( M \)-th element, which is directly related to the cumulative distribution function (CDF). Thus, efficient computation of partial sums is essential for probabilistic modeling and simulation, such as in Markov chains.

The computation of partial sums $S_M$ as given in \mref{eq_partial_sum} is non-trivial for a general $M$. We provide a novel quantum algorithm (ref.~Algorithm \ref{alg_weighted_partial_sum}) for computing $S_M$ up to a known constant normalization factor $\frac{1}{\sqrt{M}}$. The key to this is to construct a quantum circuit where the associated unitary matrix of size $N \times N$ has the following first row: \begin{align}   {\bm{\chi}} = \frac{1}{\sqrt{M}}\begin{bmatrix} 1 & 1 & \cdots & 1 & 0 & \cdots & 0 \end{bmatrix}, \end{align} with the first $M$ elements equal to $1$, followed by $N-M$ zeros. It is easy to see that 
$S_M = \sqrt{M} {\bm{\chi}} {\bm{f}}  $

If $M$ is of the form of $2^r$ for some positive integer $ r \leq n$, then the unitary matrix corresponding to $I^{\otimes (n-r)} \otimes H^{r}$ has the desired property.
Here $I$ and $H$ are $2 \times 2$ matrices associated with identity and the Hadamard transforms, respectively. Clearly, this case is trivial. However, when $M$ is not of this form (i.e., $M \neq 2^r$ for any positive integer $r$), then it is not obvious how to efficiently create  such a unitary matrix that could be used to evaluate the partial sum $S_M$. In Algorithm \ref{alg_weighted_partial_sum}, we present an efficient algorithm that constructs the desired unitary matrix with both gate complexity and circuit depth of \( O(\log_2 M) \), which can be used to evaluate the partial sum \( S_M \).

Recently, quantum and classical-quantum hybrid algorithms have been developed for solving nonlinear differential equations \cite{shukla2023hybridode, kyriienko2021solving, childs2020quantum}, optimization \cite{blekos2024review, shukla2019trajectory, sandesara2024quantum, cerezo2021variational}, as well as for image processing \cite{Shukla_Vedula_2022b, rohida2024hybrid, wang2022review} and signal processing \cite{shukla2023quantum}.
It is important to note that numerical integration is relevant for many of the above applications.
A quantum subroutine for numerical integration, which serves as a key component in these broader applications, would be highly valuable across various fields.

Many quantum techniques for numerical integration, including those based on Monte Carlo methods, are well-established in the literature \cite{herbert2022quantum, shimada2020quantum, heinrich2003quantum, abrams1999fast, shu2024general}. Our algorithm takes a distinct approach and, under the assumption that the input quantum state is either efficiently prepared or readily available from prior quantum computation, achieves a computational cost (in terms of both gate complexity and circuit depth) of \( O(\log_2 M) \), where \( M \) is the number of terms in the partial sum. The desired output (weighted partial sums) of our algorithms (Algorithm \ref{alg_weighted_partial_sum} and Algorithm \ref{alg_weighted_partial_sum_gen}) is embedded in the amplitude of the \(\ket{0}\) state, enabling further quantum processing to be performed on this state. Moreover, if a classical result is required, only the amplitude of the \(\ket{0}\) state needs to be estimated. In such cases, the computational cost of amplitude estimation (which depends on the desired precision)  \cite{kitaev1995quantum, brassard2002quantum, suzuki2020amplitude, grinko2021iterative, giurgica2022low} must be accounted for in addition to the \( O(\log_2 M) \) cost of our algorithm.
In contrast, many Monte Carlo based methods have a complexity of \( O(N) \), when \( N \) sampling points are used \cite{shu2024general}.

Next, we provide a brief outline of the remaining sections of the paper. Section \ref{sec:algoI} introduces the main quantum algorithm for computing partial sums, followed by Section \ref{sec:algoICor}, which discusses the correctness of Algorithm \ref{alg_weighted_partial_sum}. Section \ref{sec:algoII} extends the algorithm to handle weighted partial sums (ref. Algorithm \ref{alg_weighted_partial_sum_gen}), while Section \ref{sec:algoIICor} addresses the correctness of this generalization. In Section \ref{sec:Complex}, we analyze the computational complexity of both algorithms, emphasizing their efficiency. Section \ref{sec:Application} covers applications, including weighted partial sums and numerical integration, supported by computational examples. Further generalizations are discussed in Section \ref{sec:gen}. Finally, Section \ref{sec:conclusion} concludes the paper by summarizing the key findings.

\section{A quantum algorithm for computing the partial sums} \label{sec:algoI}

In this section, we will describe our algorithm for computing the partial sum $ S_M = \sum_{k=0}^{M-1} f_k $ 
corresponding to the given normalized vector
where  \( \begin{bmatrix} f_0 & f_1 & \hdots & f_{N-1} \end{bmatrix}\), and where $1 < M \leq N = 2^n$ (see \meqref{eq_partial_sum}). 
The input to our algorithm is normalized input state 
 $ \ket{f} = \sum_{s=0}^{N - 1} {f_s} \, \ket{s} $ and a positive integer $ M $  with $ 2 \leq M \leq N $. The goal of our quantum algorithm is to produce a quantum state  $\ket{\widetilde{f}} = \sum_{s=0}^{2^n - 1}  \widetilde{f_s} \, \ket{s} $, such that the desired partial sum (up to a normalization factor of $\frac{1}{\sqrt{M}}$)  is the amplitude of the computational basis state 
 $\ket{0}$, i.e., $\widetilde{f_0} = \frac{1}{\sqrt{M}} \, \sum_{k=0}^{M-1} f_k $.

As mention earlier in the introduction section, the key idea of our algorithm is to create a quantum circuit such that the corresponding unitary matrix has the first row as follows:
\begin{align} \label{eq_first_row}
    \begin{bmatrix}
        1 & 1 & \cdots & 1 & 0 & \cdots & 0
    \end{bmatrix},
\end{align}
where the first  $M$ entries are all $1$ and the rest are $0$.
It is clear that the action of such a unitary matrix on $ \ket{f} = \sum_{s=0}^{N - 1} {f_s} \, \ket{s} $ would result in the desired state $\ket{\widetilde{f}} = \sum_{s=0}^{N - 1}  \widetilde{f_s} \, \ket{s} $ with $\widetilde{f_0} = \frac{1}{\sqrt{M}} \, \sum_{k=0}^{M-1} f_k $.

If $M$ is of the form $2^r$ for a positive integer $1 \leq r \leq n$, then creating a unitary operator with the first row of its matrix as given in \meqref{eq_first_row}, is easy and it could be achieved by the application of the unitary operator $I^{\otimes 2^n - 2^r} \times H^{\otimes r}$, where $H$ is the Hadamard gate. However, when $M$ is not of the form $2^r$, then creating such a unitary matrix is non-trivial. The precise steps for constructing a quantum circuit that implements the corresponding unitary operator are outlined in Algorithm \ref{alg_weighted_partial_sum}.

\begin{algorithm}[H] \label{alg_weighted_partial_sum}
	\DontPrintSemicolon
	\KwInput{A normalized input state 
 $ \ket{f} = \sum_{s=0}^{N - 1} {f_s} \, \ket{s} $, a positive integers $ M $ and $n$,  with $ 2 \leq M \leq N = 2^n $.}
	\KwOutput{ $\ket{\widetilde{f}} = \sum_{s=0}^{N - 1}  \widetilde{f_s} \, \ket{s} $, where $\widetilde{f_0} = \frac{1}{\sqrt{M}} \, \sum_{j=0}^{M-1} f_j $.} 
	\Fn{AmplitudePartialSum($\ket{f}$, $M$, $n$)}{
        Receive the input state $\ket{f}$ on an $n$ qubit register $\mathcal{R}$, where the qubits are  labeled as $q_{n-1} \, q_{n-2} \, \ldots q_1 \, q_0$. \\
        \If {$M$ is of the form $2^r$ for a positive integer $1 \leq r \leq n$\\}
        {Apply  ($ H $) gate on  $ \ket{q_{i}} $ for $ i = 0$ to $r-1$ 
        }
        \Else{
        \tcc{ $  l_0, l_1, \ldots\,,l_k  $ is an ordered sequence of numbers representing the locations of $ 1 $ in the reverse binary representation of $ M $.}
		Compute $  l_0, l_1, \ldots\,,l_k  $, where $ M = \sum_{j=0}^{k} \, 2^{l_j} $ with $ 0 \leq l_0 < l_1 < \ldots < l_{k-1} < l_k \leq n-1 $. \\
		Set $ M_{k-1} = M- 2^{l_k} $. \\
        \For{$ m=k-1 $ down to $1$}{
        Apply a controlled Hadamard ($ H $) gate on $ \ket{q_{i}} $ for  $i= l_{m+1} -1 $, $ l_{m+1} -2 $, $ \ldots $,  $ l_m + 1$,  $ l_m$,   conditioned on $ {{q}_{l_{m+1}}} $  being equal to $ 0 $.\\
        Set $ M_{m-1} = M_{m} - 2^{l_m} $. \\
		Apply a controlled $ R_Y(\theta_m) $ gate, with  $ \theta_m =  2 \arccos \left(\sqrt {\frac{2^{l_m}}{M-M_{m-1}}}\right)$, on $ \ket{{q}_{l_{m+1}}} $  conditioned on $  q_{l_m}$ being $ 0 $. \\
	      }
       Apply a controlled Hadamard ($ H $) gate on $ \ket{q_{i}} $ for $ i= l_1 -1 $, $l_1 -2$, $ \ldots $, $ l_0 + 1$, $ l_0 $,  conditioned on $  q_{l_1}$ being equal to $ 0 $. \\
       Apply the rotation gate $ R_Y(\theta_0) $ on $ \ket{q_{l_1}}$, where  $ \theta_0 = 2 \arccos \left(\sqrt {\frac{M_0}{M}}\right)  $. \\
       \tcc{If $ M $ is an even number, then apply Hadamard gates on the rightmost $l_0$ qubits.} 
	  \If{ $ l_0 > 0  $} 
		{Apply $ H $ gate on  $ \ket{q_{i}} $ for $ i = 0$, $ 1$, $ \ldots $, $ l_0 -1 $.}
        Apply $ X $ gate on $ \ket{q_{i}} $ for $ i = l_1$, $ l_2$, $ \ldots $, $ l_k $. 
        \\
}
\tcc{The register $\mathcal{R}$ now contains the quantum state  $\ket{\widetilde{f}} = \sum_{s=0}^{N - 1}  \widetilde{f_s}$, where $\widetilde{f_0} = \frac{1}{\sqrt{M}} \, \sum_{j=0}^{M-1} f_j $.} 	
		
	\Return{the quantum state $\ket{\widetilde{f}}$ on the register $\mathcal{R}$.}
}
	\caption{A quantum algorithm for computing the partial sums of amplitudes.}
\end{algorithm}

\begin{remark}\leavevmode
\begin{enumerate}[(a)]
   \item  By removing the restrictions on $\theta_0$ and $\theta_m$ (the rotation angles for the rotation and controlled rotation gates) in lines $11$ and $12$ in Algorithm \ref{alg_weighted_partial_sum} (i.e.,  by using arbitrary  $\theta_0$ and $\theta_m$) more general weighted partial sums can be obtained (as discussed in \mref{sec:algoII}).
   \item One can also use Algortihm 1 in \cite{shukla2024efficient} to compute partial sums by preparing the uniform superposition state $\ket{\psi} = \frac{1}{\sqrt{M}}  \sum_{k=0}^{M-1} \, \ket{k}$ and then computing the inner product $\braket{\psi}{f}$ using the swap test. However, this will require double the number of qubits in comparison to using Algorithm \ref{alg_weighted_partial_sum} and amplitude estimation techniques.
   \end{enumerate}
\end{remark}

\begin{figure}[H]
  \begin{center}
       \includegraphics[width=0.47\textwidth]{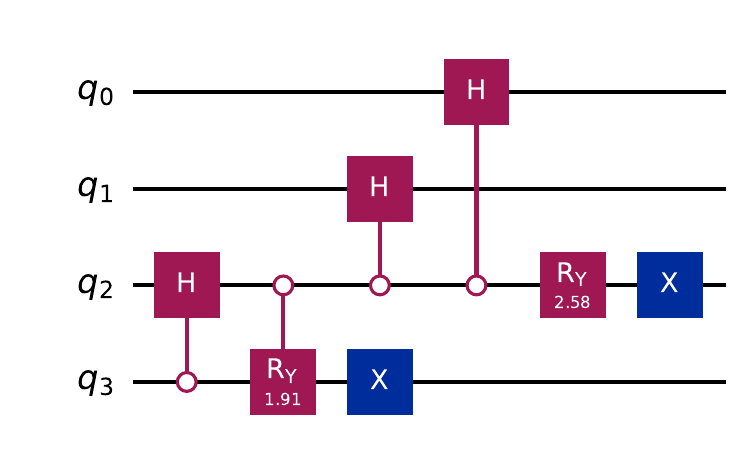}
        \includegraphics[width=0.52\textwidth]{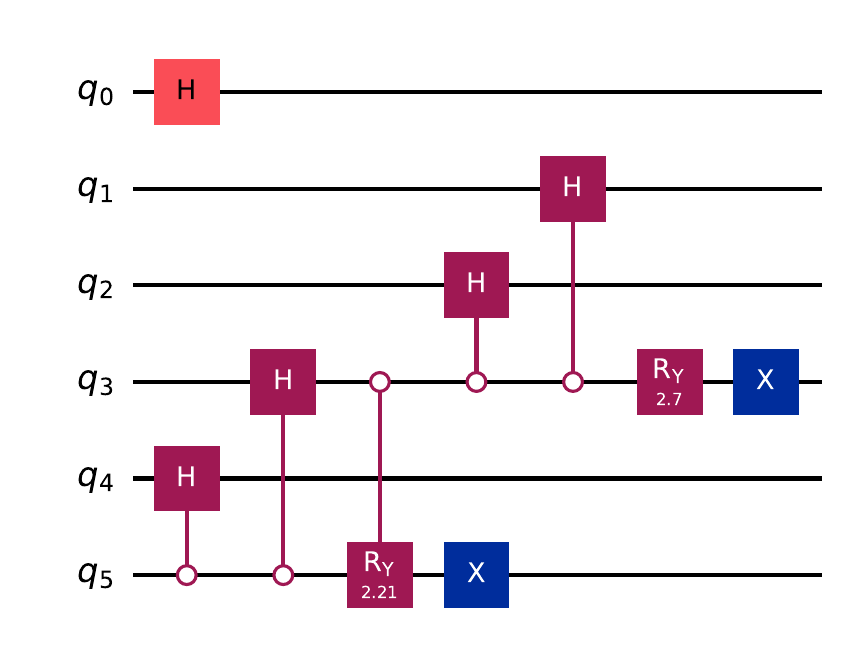}
  \end{center}
	 	\caption{Quantum circuits created by Algorithm \ref{alg_weighted_partial_sum} for computing the partial sums for $M=13$, $n=4$ (left)  and $M=42$, $n=6$ (right).}
	 	\label{fig:uniform_ex_one}
	 \end{figure}

Quantum circuits created by Algorithm \ref{alg_weighted_partial_sum} for computing the partial sums for $M=13$, $n=4$ (i.e., $N=16$) and $M=42$, $n=6$ (i.e., $N =64$) are shown in \mfig{fig:uniform_ex_one}. Some applications of Algorithm \ref{alg_weighted_partial_sum} related to evaluation of partial sums and numerical integration are provided in \mref{sec:Application}. In particular, an example that uses the quantum circuit created by Algorithm \ref{alg_weighted_partial_sum}, for $M=13$, $n=4$ (shown on the left of \mfig{fig:uniform_ex_one}), is discussed in \mref{sec:ex_one}.

 \section{Discussion on the correctness of Algorithm \ref{alg_weighted_partial_sum}}
 \label{sec:algoICor}

We now present a more detailed explanation of the steps in Algorithm \ref{alg_weighted_partial_sum} and discuss the reasoning behind its correctness. 
Since the desired partial weighted sum is produced as the amplitude $\widetilde{f_0} = \frac{1}{\sqrt{M}} \, \sum_{j=0}^{M-1} f_j $ corresponding to the $\ket{0}$ state of  $\ket{\widetilde{f}} = \sum_{s=0}^{N - 1}  \widetilde{f_s} \ket{s}$, we want to look at the first row of the unitary matrix (which we denote by $U$) corresponding to quantum circuit created by Algorithm \ref{alg_weighted_partial_sum}. If $M$ is of the form $2^r$, then the first row the unitary matrix $U$ (which corresponds to the lines $3$ and $4$ of Algorithm \ref{alg_weighted_partial_sum}) is:
\begin{align}
   \frac{1}{\sqrt{M}} \begin{bmatrix}
        1 & 1 & \cdots & 1 & 0 & \cdots & 0
    \end{bmatrix},
\end{align}
where the first $2^r$ entries are all $1$ and the rest are $0$.

Next we analyze the unitary $U$ corresponding to the quantum circuit generated by Algorithm \ref{alg_weighted_partial_sum}, when $M$ is not of the form $2^r$. This case corresponds to the lines $6$ to $16$ in Algorithm \ref{alg_weighted_partial_sum}. It is interesting to note that the unitary operator $U$ corresponding to these steps is precisely the inverse of the unitary matrix $U^{\dagger}$ corresponding to the quantum circuit generated by the efficient algorithm for preparation of uniform superposition state presented in \cite{shukla2024efficient}, i.e., Algorithm 1 in \cite{shukla2024efficient}. This can be easily verified by observing that the steps in 
Algorithm 1 in \cite{shukla2024efficient}
are in the reverse order of the steps described in lines $6$ to $16$ in Algorithm \ref{alg_weighted_partial_sum} and the rotation angles of the corresponding $R_Y$ and controlled $R_Y$ gates are negative of each other.

Since, Algorithm 1 in \cite{shukla2024efficient}
generates a uniform superposition state $\frac{1}{\sqrt{M}}  (\sum_{k=0}^{M-1} \, \ket{k}$ starting from $\ket{0}$, its unitary matrix can be expressed as:
\begin{align}
    {U^{\dagger}} = \frac{1}{\sqrt{M}} \, \begin{bmatrix}
    1      & a_{01} & a_{02} & \dots  & a_{0(N-1)} \\
    1      & a_{11} & a_{12} & \dots  & a_{1(N-1)} \\
    \vdots & \vdots & \vdots & \ddots & \vdots \\
    1      & a_{(M-1)1} & a_{(M-1)2} & \dots  & a_{(M-1)(N-1)} \\
    0      & a_{(M)1} & a_{(M)2} & \dots  & a_{(M)(N-1)} \\
    \vdots & \vdots & \vdots & \ddots & \vdots \\
    0      & a_{(N-1)1} & a_{(N-1)2} & \dots  & a_{(N-1)(N-1)} \\
\end{bmatrix},
\end{align}
where we are only concerned with the entries in the first column. Clearly, then the first row of the matrix $U$ is the first column of $U^{\dagger}$, 
\begin{align} \label{eq_def_U_M}
    U = \frac{1}{\sqrt{M}} \, \begin{bmatrix}
    1      & 1      & \dots  & 1      & 0      & \dots  & 0 \\
    \overline{a}_{01} & \overline{a}_{11} & \dots  & \overline{a}_{(M-1)1} & \overline{a}_{M1} & \dots  & \overline{a}_{(N-1)1} \\
    \overline{a}_{02} & \overline{a}_{12} & \dots  & \overline{a}_{(M-1)2} & \overline{a}_{M2} & \dots  & \overline{a}_{(N-1)2} \\
    \vdots & \vdots & \ddots & \vdots & \vdots & \ddots & \vdots \\
    \overline{a}_{0(N-1)} & \overline{a}_{1(N-1)} & \dots  & \overline{a}_{(M-1)(N-1)} & \overline{a}_{M(N-1)} & \dots  & \overline{a}_{(N-1)(N-1)}
\end{bmatrix}.
\end{align}
Finally, the action of \( U \) on the column vector \( \begin{bmatrix} f_0 & f_1 & \hdots & f_{N-1} \end{bmatrix}^T \) is as follows:
\[
\frac{1}{\sqrt{M}} \begin{bmatrix}
    1      & 1      & \dots  & 1      & 0      & \dots  & 0 \\
    \overline{a}_{01} & \overline{a}_{11} & \dots  & \overline{a}_{(M-1)1} & \overline{a}_{M1} & \dots  & \overline{a}_{(N-1)1} \\
    \overline{a}_{02} & \overline{a}_{12} & \dots  & \overline{a}_{(M-1)2} & \overline{a}_{M2} & \dots  & \overline{a}_{(N-1)2} \\
    \vdots & \vdots & \ddots & \vdots & \vdots & \ddots & \vdots \\
    \overline{a}_{0(N-1)} & \overline{a}_{1(N-1)} & \dots  & \overline{a}_{(M-1)(N-1)} & \overline{a}_{M(N-1)} & \dots  & \overline{a}_{(N-1)(N-1)}
\end{bmatrix}
\begin{bmatrix}
    f_0 \\
    f_1 \\
    \vdots \\
    f_{N-1}
\end{bmatrix}
=
\begin{bmatrix}
    \widetilde{f}_0 \\
    \widetilde{f}_1 \\
    \vdots \\
    \widetilde{f}_{N-1}
\end{bmatrix},
\]
with $\widetilde{f_0} = \frac{1}{\sqrt{M}} \, \sum_{j=0}^{M-1} f_j $, as desired.

\section{Further generalizations and comutation of weighted partial sums}
\label{sec:algoII}

In this section, we provide an algorithm (Algorithm \ref{alg_weighted_partial_sum_gen}) for  computation of a generalized form of weighted partial sums. The input state \( \ket{f} \) is expressed in the standard computational basis as 
$\ket{f} = \sum_{s=0}^{N - 1} {f_s} \, \ket{s}$,
where \( f_s \) are the coefficients of the state. The algorithm provides the following normalized quantum state as output 
$\ket{\widetilde{f}} = \sum_{s=0}^{N - 1} \widetilde{f_s} \, \ket{s};$
such that \( \widetilde{f_0} \) is the weighted sum expression defined in Eq.~(\ref{eq_gen_weight_sum}). The details of the algorithm are given below. 

\begin{algorithm}[H] \label{alg_weighted_partial_sum_gen}
	\DontPrintSemicolon
	\KwInput{A normalized input state 
 $ \ket{f} = \sum_{s=0}^{N - 1} {f_s} \, \ket{s} $, positive integer $ M $ and $n$  with $ 2 < M < N = 2^n $ and $M$ is not of the form $2^r$ for any integer $r$, and the vector ${\bm{b}} = [b_0, b_1, \ldots b_{k-1} ] $, where $k+1$ represents the sum of bits in the binary representation of $M$.}
	\KwOutput{ $\ket{\widetilde{f}} = \sum_{s=0}^{N - 1}  \widetilde{f_s} \, \ket{s} $, where $\widetilde{f_0}$ is given in \meqref{eq_gen_weight_sum}.} 
	\Fn{AmplitudePartialSum($\ket{f}$, $M$, $n$, $\bm{b}$)}{
        Receive the input state $\ket{f}$ on an $n$ qubit register $\mathcal{R}$, where the qubits are  labeled as $q_{n-1} \, q_{n-2} \, \ldots q_1 \, q_0$. \\
        \tcc{ $  l_0, l_1, \ldots\,,l_k  $ is an ordered sequence of numbers representing the locations of $ 1 $ in the reverse binary representation of $ M $.}
		Compute $  l_0, l_1, \ldots\,,l_k  $, where $ M = \sum_{j=0}^{k} \, 2^{l_j} $ with $ 0 \leq l_0 < l_1 < \ldots < l_{k-1} < l_k \leq n-1 $. \\
		Set $ M_{k-1} = M- 2^{l_k} $. \\
        \For{$ m=k-1 $ down to $1$}{
        Apply a controlled Hadamard ($ H $) gate on $ \ket{q_{i}} $ for  $i= l_{m+1} -1 $, $ l_{m+1} -2 $, $ \ldots $,  $ l_m + 1$,  $ l_m$,   conditioned on $ {{q}_{l_{m+1}}} $  being equal to $ 0 $.\\
        Set $ M_{m-1} = M_{m} - 2^{l_m} $. \\
		Apply a controlled $ R_Y(\theta_m) $ gate, where  $ \theta_m = 2 \arccos\lob{b_m}$, on $ \ket{{q}_{l_{m+1}}} $  conditioned on $  q_{l_m}$ being $ 0 $. \\
	      }
       Apply a controlled Hadamard ($ H $) gate on $ \ket{q_{i}} $ for $ i= l_1 -1 $, $l_1 -2$, $ \ldots $, $ l_0 + 1$, $ l_0 $,  conditioned on $  q_{l_1}$ being equal to $ 0 $. \\
       Apply the rotation gate $ R_Y(\theta_0) $ on $ \ket{q_{l_1}}$, with a chosen rotation angle $ \theta_0 = 2 \arccos\lob{b_0} $. \\
       \tcc{If $ M $ is an even number, then apply Hadamard gates on the rightmost $l_0$ qubits.} 
	  \If{ $ l_0 > 0  $} 
		{Apply $ H $ gate on  $ \ket{q_{i}} $ for $ i = 0$, $ 1$, $ \ldots $, $ l_0 -1 $.}
        Apply $ X $ gate on $ \ket{q_{i}} $ for $ i = l_1$, $ l_2$, $ \ldots $, $ l_k $. 
        \\
\tcc{The register $\mathcal{R}$ now contains the quantum state  $\ket{\widetilde{f}} = \sum_{s=0}^{N - 1}  \widetilde{f_s}$, where $\widetilde{f_0} $ is defined in \meqref{eq_gen_weight_sum}.} 	
		
	\Return{the quantum state $\ket{\widetilde{f}}$ on the register $\mathcal{R}$.}
}
	\caption{A quantum algorithm for computing certain weighted partial sums of amplitudes.}
\end{algorithm}

\begin{remark}
    If we impose the restriction 
    $b_m = \sqrt {\frac{2^{l_m}}{M-M_{m-1}}}$
    in the step $11$ and
    $ b_0 = \sqrt {\frac{M_0}{M}}  $ 
    in the step $13$ of Algorithm \ref{alg_weighted_partial_sum_gen}, then we reduce to the case of Algorithm \ref{alg_weighted_partial_sum} where $M$ is not of the form $2^r$ for any integer $r$.
\end{remark}
Let 
\begin{equation} \label{eq_S_r}
S_r :=
\begin{cases}
    \sum_{s = k- r}^{k} 2^{l_s} & \text{if } 0 \leq r \leq k, \\
    0 & \text{otherwise}.
\end{cases}
\end{equation}
Let us further define 
\begin{align} \label{eq_gen_weight_sum}
	\ket{\widetilde{f}_0} = 
	\frac{b_0}{\sqrt{2^{l_0}}}  \sum_{j = S_{k-1} }^{S_{k} -1}  f_j +  \frac{a_0 b_1 }{\sqrt{2^{l_1}}} \sum_{j = S_{k-2} }^{S_{k-1} -1} f_j
	+ \frac{a_0 a_1 b_2 }{\sqrt{2^{l_2}}}   \sum_{j = S_{k-3}}^{S_{k-2} - 1} f_j +
	\cdots 
	+ \frac{a_0 a_1 \ldots a_{k-2} b_{k-1}}{\sqrt{2^{l_{k-1}}}} \sum_{j= S_0 }^{S_1 - 1} f_j
	+ \frac{a_0 a_1 \ldots a_{k-2} a_{k-1} }{\sqrt{2^{l_{k}}}}\sum_{j=0}^{S_0 - 1} f_j,
\end{align}

where the coefficient  $a_r$ and $b_r$ are arbitrary real numbers satisfying the normalization requirement $\lvert a_r \rvert^2 + \lvert b_r \rvert^2 =1 $, for $r=0$ to $r=k-1$. 
Alternatively, one can write \eqref{eq_gen_weight_sum}  as

\begin{align}
	\ket{\widetilde{f}_0}  = 	\sum_{r=0}^{k} \, \gamma_{k-r} \lob{\sum_{j = S_{r-1}}^{S_r - 1} f_{  j}},
\end{align}

where \begin{align} \label{eq_short_def}
	\gamma_r  =
	\begin{cases}
		&  \frac{b_0}{\sqrt{2^{l_0}}} \quad \quad \quad \text{if } r=0, \\
		& \frac{a_0 a_1 \ldots a_{r-1} b_{r}}{\sqrt{2^{l_{r}}}} \quad \text{if }  0 <  r \leq k-1,  \\ 
		& \frac{a_0 a_1 \ldots a_{k-2} a_{k-1}} {\sqrt{2^{l_{k}}}} \quad \text{if }  r = k.
	\end{cases}
\end{align} 

We note that the coefficients \(\gamma_{k-r}\) represent the weights for the functions \(f_{j} \) for \(j = S_{r-1}\) to \(S_{r} - 1\).

\section{Discussion on the correctness of Algorithm \ref{alg_weighted_partial_sum_gen}}
\label{sec:algoIICor}

In this section,  we present a more detailed explanation of the steps in Algorithm \ref{alg_weighted_partial_sum_gen} and discuss the reasoning behind its correctness. 
Since the desired partial weighted sum is produced as the amplitude $\widetilde{f_0} = \frac{1}{\sqrt{M}} \, \sum_{j=0}^{M-1} f_j $ corresponding to the $\ket{0}$ state of  $\ket{\widetilde{f}} = \sum_{s=0}^{N - 1}  \widetilde{f_s} \ket{s}$, we want to look at the first row of the unitary matrix (which we denote by $U$) corresponding to quantum circuit created by Algorithm \ref{alg_weighted_partial_sum_gen}.
It is straightforward to verify that the unitary operator \( U \) resulting from these steps is exactly the inverse of the unitary matrix \( U^{\dagger} \) associated with the quantum circuit generated by Algorithm 1 in \cite{shukla2024efficient}, provided the restrictions on the rotation angles \(\theta_0\) and \(\theta_m\) (used in the rotation and controlled rotation gates) in lines 8 and 11 of the algorithm are removed.
The generalization of Algorithm 1 from \cite{shukla2024efficient} is discussed in Section 3 of the same work, which describes the construction of nonuniform superpositions.

At the end of Algorithm 1 in \cite{shukla2024efficient}, the quantum state obtained is  
\begin{align} \label{eq_general_expression}
	\ket{\psi_{k-1}} = 
	\frac{b_0}{\sqrt{2^{l_0}}}  \sum_{j=0}^{2^{l_0} - 1} &\ket{j + M - 2^{l_0}} +  \frac{a_0 b_1 }{\sqrt{2^{l_1}}}  \sum_{j=0}^{2^{l_1} - 1} \ket{j + M - 2^{l_0} - 2^{l_1}}
	+ \frac{a_0 a_1 b_2 }{\sqrt{2^{l_2}}}   \sum_{j=0}^{2^{l_2} - 1} \ket{j + M - 2^{l_0} - 2^{l_1} - 2^{l_2} } \nonumber\\
	\cdots \cdots
	&+ \frac{a_0 a_1 \ldots a_{k-2} b_{k-1}}{\sqrt{2^{l_{k-1}}}} \sum_{j=0}^{2^{l_{k-1}} - 1} \ket{j + M - \sum_{s=0}^{k-1} \, 2^{l_s} }
	+ \frac{a_0 a_1 \ldots a_{k-2} a_{k-1} }{\sqrt{2^{l_{k}}}}\sum_{j=0}^{2^{l_{k}} - 1} \ket{  j + M - \sum_{s=0}^{k} \, 2^{l_s}}.
\end{align}
This can be rewritten as
\begin{align} \label{eq_general_expression}
	\ket{\psi_{k-1}} = 
	\frac{b_0}{\sqrt{2^{l_0}}}  \sum_{j = S_{k-1} }^{S_{k} -1}  \ket{j} +  \frac{a_0 b_1 }{\sqrt{2^{l_1}}} \sum_{j = S_{k-2} }^{S_{k-1} -1} \ket{j }
	+ \frac{a_0 a_1 b_2 }{\sqrt{2^{l_2}}}   \sum_{j = S_{k-3}}^{S_{k-2} - 1} \ket{j } \nonumber
	\cdots 
	+ \frac{a_0 a_1 \ldots a_{k-2} b_{k-1}}{\sqrt{2^{l_{k-1}}}} \sum_{j= S_0 }^{S_1 - 1} \ket{j}
	+ \frac{a_0 a_1 \ldots a_{k-2} a_{k-1} }{\sqrt{2^{l_{k}}}}\sum_{j=0}^{S_0 - 1} \ket{  j},
\end{align}
where $S_r$ is as defined in \meqref{eq_S_r}.
Let us denote the first column of the unitary matrix $U^{\dagger}$ corresponding to the quantum circuit generated by Algorithm 1 in \cite{shukla2024efficient} in the general nonuniform case (see Section 3 of \cite{shukla2024efficient}) by ${\bm{v}}$.
It can be observed that the first $S_0$ components of ${\bm{v}}$ are $\gamma_k$, followed by $S_1 - S_0$ components of $\gamma_{k-1}$, then $S_2 - S_1$ components of $\gamma_{k-2}$, and so forth.
It then follows that
\[
{\bm{v}} = 
\begin{bmatrix}
    v_0 & v_1 & \ldots & v_j &  \ldots  & v_{N -1}
\end{bmatrix}^T,
\]
where
\begin{align} 
	v_j  =
	\begin{cases}
		& \gamma_{k-r} \quad \quad \text{if }  S_{r-1} \leq j \leq S_r -1, \quad \text{for $r=0$ to $r=k$},\\
  &  0 \quad \quad \quad \text{ if } 
 M \leq  j \leq N - 1.
	\end{cases}
\end{align} 
Then the first row of the unitary matrix $U$ (corresponding to quantum circuit created by Algorithm \ref{alg_weighted_partial_sum_gen}) is given by ${\bm{v}}^\dagger = {\bm{v}}^T $. Further, the action of \( U \) on the column vector \( {\bm{f}} = \begin{bmatrix} f_0 & f_1 & \hdots & f_{N-1} \end{bmatrix}^T \) results in 
$ {\bm{\widetilde{f}}}  =  \begin{bmatrix} \widetilde{f}_0 & \widetilde{f}_1 & \hdots & \widetilde{f}_{N-1} \end{bmatrix}^T $, 
where  $\widetilde{f_0} =  {\bm{v}}^T {\bm{f}}$ which is the same is given in \meqref{eq_gen_weight_sum}.

\section{Computational complexity}
\label{sec:Complex}

 Let \( M = \sum_{j=0}^{k} 2^{l_j} \) with \( 0 \leq l_0 < l_1 < \ldots < l_{k-1} < l_k \leq n-1 \), where $N=2^n$. For $M$ given as above, Algorithms \ref{alg_weighted_partial_sum} and \ref{alg_weighted_partial_sum_gen} 
require \( l_k + 2k \) quantum gates, including one rotation gate \( R_Y(\theta) \), \( k \) Pauli-\(X\) gates, \( l_0 \) Hadamard gates (if \( l_0 > 0 \)), \( l_k - l_0 \) controlled-Hadamard gates, and \( k - 1 \) controlled rotation \( R_Y(\theta) \) gates.

For cases where \( M = 2^r \) for some \( r \in \mathbb{N} \), Algorithms \ref{alg_weighted_partial_sum} and \ref{alg_weighted_partial_sum_gen} 
require \( r \) Hadamard gates. Thus, the number of elementary gates required for Algorithms \ref{alg_weighted_partial_sum} and \ref{alg_weighted_partial_sum_gen} are
 \( O(\log_2 M) \). The circuit depth is also  \( O(\log_2 M) \), while the number of qubits needed is $n =  O(\log_2 N)$.

We note that Algorithms \ref{alg_weighted_partial_sum} and \ref{alg_weighted_partial_sum_gen} generate the partial sums as coefficients of the \( \ket{0} \) state. Further quantum processing can be carried out on this state by using the appropriate control gates. If classical results are required, the partial sums can be extracted from the quantum state by  amplitude estimation techniques \cite{kitaev1995quantum, brassard2002quantum, suzuki2020amplitude, grinko2021iterative, giurgica2022low}.

\section{Applications}
\label{sec:Application}

We assume that we have access to the function \(\mathcal{S}_M(\ket{v})\) that computes the partial sum of the first \(M\) components of the normalized vector \(\ket{v}\) of length \(N = 2^n\), scaled by \(\frac{1}{\sqrt{M}}\) using Algorithm \ref{alg_weighted_partial_sum}. Specifically, we have:

\[
\mathcal{S}_M(\ket{v}) = \frac{1}{\sqrt{M}} \left( v_0 + v_1 + \dots + v_{M-1} \right),
\]
where \(M\) is the number of terms summed.

We assume that the quantum state \(\ket{v}\) is already available. The preparation of such a state involves state-preparation techniques, which are outside the scope of this work. For our purposes, we consider the state to be either prepared beforehand or produced as the result of a preceding quantum algorithm.

\subsection{Weighted partial sums and numerical integration}
Next, we will provide computational examples for evaluation of partial sums and numerical integration.

\subsubsection{Computational example: partial sums} \label{sec:ex_one}

\begin{figure}
    \centering
\includegraphics[width=0.6\textwidth]{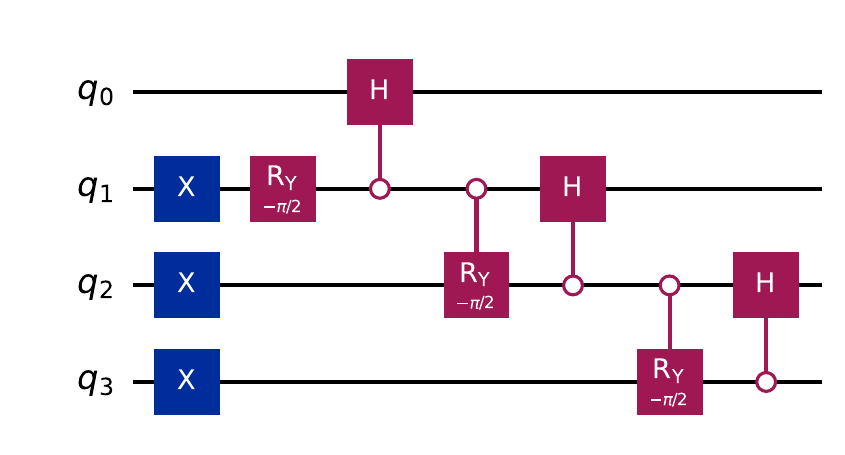}
    \caption{Quantum circuit for creating the state $\ket{v} = \frac{1}{\sqrt{64}} \left( \ket{0} + \ket{1} + \ket{2} + \ket{3} + \ket{4} + \ket{5} + \ket{6} + \ket{7}    \right) +  	\frac{1}{\sqrt{32}} \left( \ket{8} + \ket{9} + \ket{10} + \ket{11} \right)  +  	\frac{1}{\sqrt{8}} \left( \ket{12} + \ket{13} \right) +
\frac{1}{\sqrt{2}} \left( \ket{14} \right).$}
    \label{fig:nu}
\end{figure}

Let us consider a normalized vector \(\ket{v}\) where the length \(N = 2^n\). In this example, let \(N = 16\), i.e., \(n = 4\), and the vector \(\ket{v}\) is given as:
\begin{align*}
\ket{v} = \sum_{k=0}^{15} v_k \,\ket{k} = &\frac{1}{\sqrt{64}} \left( \ket{0} + \ket{1} + \ket{2} + \ket{3} + \ket{4} + \ket{5} + \ket{6} + \ket{7}    \right) +  	\frac{1}{\sqrt{32}} \left( \ket{8} + \ket{9} + \ket{10} + \ket{11} \right)  \\ & +  	\frac{1}{\sqrt{8}} \left( \ket{12} + \ket{13} \right) +
\frac{1}{\sqrt{2}} \left( \ket{14} \right). 
\end{align*}
We note that the state $\ket{v}$ can be prepared using the quantum circuit shown in \mfig{fig:nu}, where the input $\ket{q_3 q_2 q_1 q_0}$ is set to $ \ket{0000}$.

\begin{figure}[H]
    \centering
\includegraphics[width=0.4\textwidth]{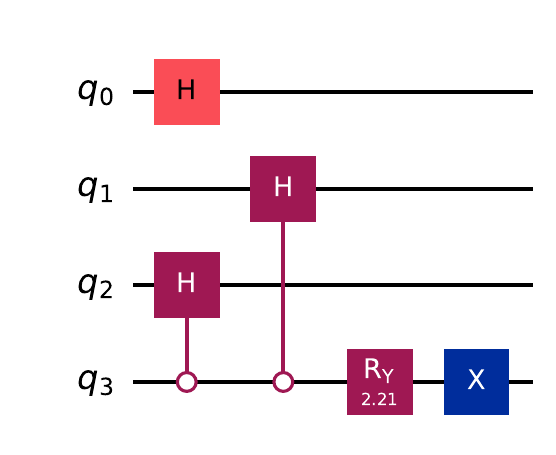}
    \caption{ Quantum circuits created by Algorithm 1 for computing the partial sums for $M = 10$.}
    \label{fig:inux}
\end{figure}

Given the quantum state $\ket{v}$,  
one can compute partial sums for different values of \(M\), where \(2 \leq M \leq 15\). 
For example, for $M=10$, the quantum circuit shown in \mfig{fig:inux}, with the input $\ket{v}$, creates the quantum state 
$\sum_{s=0}^{15} \, c_s \, \ket{s}$ such that the 
$c_0 =  \frac{1}{\sqrt{10}} \sum_{k=0}^{9} v_k  =  \frac{1}{\sqrt{10}} \lob{1 + \frac{1}{\sqrt{8}}} $.
Similarly, for $M=13$, the 
the quantum circuit shown in \mfig{fig:uniform_ex_one} (on the left), with the input $\ket{v}$, produces the state 
$\sum_{s=0}^{15} \, c_s \, \ket{s},$ such that 
$c_0 =  \frac{1}{\sqrt{13}} \sum_{k=0}^{12} v_k  = \frac{1}{\sqrt{13}} \lob{1 + \frac{1}{\sqrt{2}} + \frac{1}{\sqrt{8}}}$.

\subsubsection{Computational example: numerical integration}

The computation of weighted partial sums is also related to numerical integration. In the following, we provide an example involving numerical integration.

Consider the function \(f(x) = \sin(\pi x)\) on the interval \([0, 1]\), which we sample at the midpoints of \(N = 16\) equally spaced subintervals. The midpoints of these intervals are given by \(x_k = \frac{2k+1}{2N}\) for \(k = 0, 1, \dots, N-1\). The sampled values form the vector:
\[
{\bm{\widetilde{v}}} = \begin{bmatrix} \sin\left( \frac{\pi}{32} \right) & \sin\left( \frac{3\pi}{32} \right) & \sin\left( \frac{5\pi}{32} \right) & \cdots \cdots & \sin\left( \frac{31\pi}{32} \right) \end{bmatrix}.
\]
The vector \({\bm{\widetilde{v}}}\) is normalized by dividing by its \(L_2\)-norm \(\|{\bm{\widetilde{v}}}\|_2    \) to obtain the normalized vector \({\bm{v}}\):
\[
{\bm{v}} = \frac{{\bm{\widetilde{v}}}}{\|{\bm{\widetilde{v}}}\|_2}.
\]
We assume that the normalized vector \({\bm{v}}\) is available as a quantum state
\[
\ket{v} = \sum_{k=0}^{N-1} v_k \ket{k},
\]
where the amplitudes \(v_k\) are the components of the vector \({\bm{v}}\). As noted earlier, we assume the state is either prepared in advance or is the output of a previous quantum computation. 
For \(M = 12\), corresponding to \(a = \frac{M}{N} = 0.75\), we approximate the integral:
\[
\int_0^{a} \sin(\pi x) \, dx = \int_0^{0.75} \sin(\pi x) \, dx \approx \mathcal{S}_{12}(\ket{v}) \times \sqrt{12} \approx 0.5442628374252914 .
\]

The computational complexity of our algorithm is \( O(\log_2 N) \), owing to the \( O(\log_2 N) \) complexity of Algorithm \ref{alg_weighted_partial_sum}, while ignoring the state preparation cost (or assuming that the input quantum state is available from prior quantum computation steps). The integration result appears as the coefficient of the \( \ket{0} \) state, which can be processed further in the next step via a quantum algorithm. As mentioned earlier, if needed, one can extract the classical result through amplitude estimation techniques. 

We note that several quantum techniques for numerical integration, such as those based on Monte Carlo approaches, exist in the literature. The computational costs of many quantum algorithms for numerical integration are listed in Table 1 of \cite{shu2024general}. Our algorithm offers a distinct approach, and under the previously mentioned assumptions on state preparation, its computational cost \( O(\log_2 M) \) compares favorably to other Monte Carlo based methods (which have the complexity of  \( O(N) \), if \( N \) points are used for Monte Carlo sampling).

There are many applications that involve the computation of weighted partial sums, such as calculating cumulative probabilities in statistics, determining moving averages in time series analysis, evaluating risk and portfolio values in finance, and analyzing signal processing in engineering.

\section{Some other generalizations}
\label{sec:gen}

Algorithms \ref{alg_weighted_partial_sum} and \ref{alg_weighted_partial_sum_gen} can be applied to a subset of input qubit register to obtain further generalizations and interesting results on partial sums. For instance, suppose one is interested in computing only the even or odd partial sums
for the vector  
$\begin{bmatrix} f_0 & f_1 & \hdots & f_{2N-1} \end{bmatrix}$, i.e., $S_E = \sum_{k=0}^{M} \, f_{2k}$ or
or $ S_O = \sum_{k=0}^{M} \, f_{2k+1}$, where $ M \leq N$. 
It is clear that,  
$\bra{0} U \otimes I \ket{f} = S_E$ and $\bra{0} U \otimes X \ket{f} = S_O$, 
where $U$ is the unitary matrix defined in \meqref{eq_def_U_M} corresponding to the quantum circuit created by Algorithm \ref{alg_weighted_partial_sum}, $\ket{f} = \sum_{s=0}^{2 N - 1} f_s \, \ket{s}$, and $X = \mat{0}{1}{1}{0}$ and $I = \mat{1}{0}{0}{1}$ are unitary matrices corresponding to the Pauli $X$ gate and the identity gate, respectively.
If we replace $X$ or $I$ in the above by a more general $2^r \times 2^r $ unitary matrix $V$ (which we assume can be prepared efficiently), then 
$$\bra{0} U \otimes V \ket{g} = \sum_{k=0}^{2^r M} v_{\lob{k \mod 2^r}} \,  g_k,$$
where
$\ket{g} = \sum_{s=0}^{2^r N - 1} g_s \, \ket{s}$ and the first row of $V$ is $
\begin{bmatrix}
    v_0 & v_1 & \ldots & v_j &  \ldots  & v_{2^r -1}
\end{bmatrix}$. The above approach can further be generalized by considering 
the unitary of the form  $W \otimes S_L U S_R \otimes V$, where we assume that $W$, $S_L$, $S_R$ and $V$ can be prepared efficiently.
It is clear that one can compute a large class of weighted partial sums efficiently using the approach described above.

\section{Conclusion}
\label{sec:conclusion}
This work introduces an efficient quantum algorithm for computing partial sums and weighted partial sums.
 Given a normalized vector \( {\bm{f}} = \begin{bmatrix} f_0 & f_1 & \hdots & f_{N-1} \end{bmatrix} \) where \( N = 2^n \), the goal is to compute the partial sum \( S_M = \sum_{k=0}^{M-1} f_k \) for any integer \( M \leq N \). 
This problem has many applications including numerical integration, cumulative probability distributions, and probabilistic modeling. The proposed method uses a custom unitary construction to evaluate partial sums of a given normalized vector (up to a known constant normalization factor), achieving gate complexity and circuit depth of \(O(\log_2 M)\), where $M$ is an arbitrary positive integer denoting the number of terms in the partial sum. 
We also presented further generalizations that allow for evaluation of partial sums of even or odd components or more complex weighted sums across defined intervals.

\section{Declarations}
\paragraph{Data availability statement:}
Data sharing is not applicable to this article as no datasets were generated or analyzed during the current study.	

\paragraph{Competing interests statement:}	
The authors have no competing interests to declare that are relevant to the content of this article.


\end{document}